\documentclass{IEEEtran4PSCC}
\ifCLASSINFOpdf
\else
\fi
\usepackage[cmex10]{amsmath}
\usepackage{mathtools}
\usepackage{breqn}
\usepackage{tabularx}
\usepackage{tabu}
\usepackage{booktabs}
\usepackage{tikz}
\usepackage{subfig}
\usetikzlibrary{shapes,arrows,positioning,calc}
\usepackage[noadjust]{cite}
\usepackage{csquotes}
\usepackage{dirtytalk}
\usepackage{amsfonts} 
\usetikzlibrary{decorations.pathmorphing}

\DeclareFontFamily{U}{mathb}{\hyphenchar\font45}
\DeclareFontShape{U}{mathb}{m}{n}{<5> <6> <7> <8> <9> <10> gen * mathb
<10.95> mathb10 <12> <14.4> <17.28> <20.74> <24.88> mathb12}{}
\DeclareSymbolFont{mathb}{U}{mathb}{m}{n}

\DeclareMathSymbol{\rcirclearrow}{\mathbin}{mathb}{'367}
\makeatletter
\newcommand*{\Rcirclearrow}{\mathpalette\@Rcirclearrow{}}
\newcommand*{\@Rcirclearrow}[1]{%
    \mathbin{\ooalign{\hphantom{$#1\rcirclearrow$}\cr\hss\raisebox{0.05ex}{%
                \scalebox{0.9}{%
                    \rotatebox[origin=c]{270}{%
                        $#1\rcirclearrow$}}}\hss}}}
\makeatother

\begin{document}
%
\title{Optimized Control of Variable Speed Hydropower for Provision of Fast Frequency Reserves}

\tikzset{
block/.style = {draw, fill=white, rectangle, minimum height=2.5em, minimum width=2.5em},
block2/.style = {draw, fill=white, rectangle, minimum height=2cm, minimum width=1cm},
block3/.style = {fill=white, rectangle, minimum height=0.5cm, minimum width=0.5cm},
lblock/.style = {draw, fill=white, rectangle, minimum height=8em, minimum width=3em},
tmp/.style  = {coordinate}, 
point/.style  = {coordinate},
sum/.style= {draw, fill=white, circle, node distance=1cm},
input/.style = {coordinate},
output/.style= {coordinate},
pinstyle/.style = {pin edge={to-,thin,black}
}
}
\tikzset{snake it/.style={decorate, decoration=snake}}

\author{
\IEEEauthorblockN{Tor Inge Reigstad\\ Kjetil Uhlen}
\IEEEauthorblockA{Department for Electric Power Engineering \\
Norwegian University of Science and Technology (NTNU)\\
Trondheim, Norway\\
\{tor.inge.reigstad, kjetil.uhlen\}@ntnu.no}
}


\maketitle

\begin{abstract}
This paper deals with the design of controllers for variable speed hydropower (VSHP) plants with the objective of optimize the plants’ performance. The control objectives imply enabling fast responses to frequency deviations while keeping the electric and hydraulic variables within their constraints. A model predictive controller (MPC) was developed to coordinate the turbine controller with the virtual synchronous generator (VSG) control of the power electronics converter. The simulation results show that the VSG is able to deliver fast power responses by utilizing the rotational energy of the turbine and the generator. The MPC controls the guide vane opening of the turbine to regain the nominal turbine rotational speed. If this is not possible due to the constraints of the hydraulic system, the MPC adjusts the power output of the VSHP by changing the VSG power reference. The proposed control system allows the VSHP to provide fast frequency reserves (FFR).
\end{abstract}


\thanksto{This work was supported by the Research Council of Norway under Grant 257588 and by the Norwegian Research Centre for Hydropower Technology (HydroCen).}

\section{Introduction}

Variable speed operation of hydropower plants is currently being investigated, and is motivated by several factors. One key factor is the potential for providing ancillary services, such as fast frequency reserves (FFR). More renewables like wind and solar energy increase the need for flexible production and loads to balance the grid and maintain the power system security. Variable speed hydropower (VSHP) may provide this flexibility with virtual inertia (VI) control by utilizing the rotational energy of the turbine and the generator, both in production and in pumping mode. Challenges and opportunities for VSHP are further explained in \cite{valavi2016variable}. The hypothesis is that the VSHP can offer additional ancillary services, contributing to improving frequency control and maintaining grid stability, thus allowing for higher penetration of variable renewables in the grid. Complete utilization of this potential comprises the development of an advanced control system optimizing the operation of the power plant while considering the constraints in the electric and the hydraulic systems. This can be achieved by combining VI control for improving the power response to frequency deviations with model predictive control (MPC) for handling the internal control of the VSHP.

Research on the use of MPC for control of hydropower plants and frequency control is limited, however, both locally and centralized based MPCs are used for this purpose. In \cite{beus2018application}, a local MPC controller is used for hydro turbine governor control in a conventional power plant. The Francis turbine is represented by a linearized hygov-model, the guide vane opening speed is limited and generalized predictive control is used to solve the optimization problem. MPC is also used for frequency control as in \cite{elsisi2018improving}. A bat-inspired algorithm is utilized to optimize the MPC design for load frequency control of superconducting magnetic storage and capacitive energy storage.

A centralized MPC considering limitations on tie-line power flow, generation capacity, and generation rate of change is studied for load frequency control in \cite{ersdal2013applying,ersdal2016model,ersdal2016model2}, applying both linear and nonlinear MPC. MPC can also be used to damp oscillations in the AC system by minimizing the generators’ frequency deviation from the average system frequency by a global MPC-based grid control \cite{fuchs2014stabilization,sanz2017effective, azad2013damping, jain2015model}. This control layout can be modified to also control voltage and ensure voltage stability \cite{imhof2014voltage}.

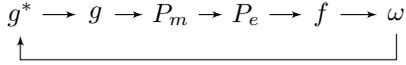
\begin{figure}
    \centering
    \begin{tikzpicture}[auto, node distance=2.5cm,>=latex']
    
    \node [block3] (gref) {$g^*$};
    \node [block3, right of=gref, node distance = 1.0cm] (g) {$g$};
    \draw [->] (gref) --  (g);
    \node [block3, right of=g, node distance = 1.0cm] (Pm) {$P_m$};
    \draw [->] (g) --  (Pm);
    \node [block3, right of=Pm, node distance = 1.0cm] (Pe) {$P_e$};
    \draw [->] (Pm) --  (Pe);
    \node [block3, right of=Pe, node distance = 1.0cm] (f) {$f$};
    \draw [->] (Pe) --  (f);
    \node [block3, right of=f, node distance = 1.0cm] (omega) {$\omega$};
    \draw [->] (f) --  (omega);
    
    \node [point, below of =omega, node distance = 0.6cm] (omegab) {};
    \draw [->] (omega) --  (omegab) -| (gref);

\end{tikzpicture}
\caption{Control layout of conventional hydropower plant} \label{fig:con}
\end{figure}

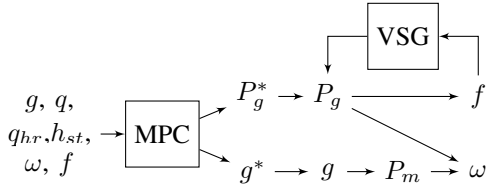
\begin{figure}
    \centering
    \begin{tikzpicture}[auto, node distance=2.5cm,>=latex']
    
    \node [block] (mpc) {MPC};
    \node [point, above of =mpc, node distance = 0.5cm] (mpca) {};
    \node [point, below of =mpc, node distance = 0.5cm] (mpcb) {};
    
    \node [block3, right of=mpcb, node distance = 1.2cm] (gref) {$g^*$};
    \draw [->] (mpc) --  (gref);
    \node [block3, right of=gref, node distance = 1.0cm] (g) {$g$};
    \draw [->] (gref) --  (g);
    \node [block3, right of=g, node distance = 1.0cm] (Pm) {$P_m$};
    \draw [->] (g) --  (Pm);
    \node [block3, right of=Pm, node distance = 1.0cm] (omega) {$\omega$};
    \draw [->] (Pm) --  (omega);
    
    \node [block3, right of=mpca, node distance = 1.2cm] (Pgref) {$P_g^*$};
    \draw [->] (mpc) --  (Pgref);
    \node [block3, right of=Pgref, node distance = 1.0cm] (Pg) {$P_g$};
    \draw [->] (Pgref) --  (Pg);
    \node [block3, right of=Pg, node distance = 2.0cm] (f) {$f$};
    \draw [->] (Pg) --  (f);
    
    \draw [->] (Pg) --  (omega);
    \node [point, above of =f, node distance = 0.8cm] (fa) {};
    \node [block, left of=fa, node distance = 1.0cm] (VSG) {VSG};
    \draw [->] (f) |-  (VSG);
    \draw [->] (VSG) -|  (Pg);

    \node [block3, left of=mpc, node distance = 1.5cm] (in) {$q_{hr}$,$h_{st}$,};
    \node [block3, above of =in, node distance = 0.4cm] (mpca) {$g$, $q$, };
    \node [block3, below of =in, node distance = 0.4cm] (mpcb) { $\omega$, $f$};
    \draw [->] (in) --  (mpc);
    
    
    
\end{tikzpicture}
\caption{Control layout of VSHP plant with MPC control} \label{fig:vshp}
\end{figure}

A PID controller is utilized to control the guide vane opening of a VSHP in \cite{reigstad2019variable} while virtual inertia control methods for VSHP are investigated in \cite{reigstad2019virtual}. The internal control of the VSHP and the virtual inertia control is not coordinated and a more advanced controls system is needed to ensure that the power response of the virtual control will not cause problems for the internal control of the power plant. In this paper, the VSHP control is improved by proposing a new control scheme: MPC and virtual synchronous generator (VSG) control are combined to optimize the frequency response of the power plant while keeping the electric and hydraulic variables within their limits. While a conventional hydropower plant has a direct relation between guide vane opening reference $g^*$, guide vane opening $g$, mechanical power $P_m$, electrical power $P_e$, frequency $f$ and turbine rotational speed $\omega$ as shown in Figure \ref{fig:con}, the VSHP enables one more degree of freedom to control power and speed. The proposed control scheme utilizes the VSHP output power $P_g$ to control the frequency $f$ while the guide vane opening reference $g^*$ and the VSHP output power reference $P_g^*$ control the turbine rotational speed $\omega$, as indicated in Figure \ref{fig:vshp}. There is still a direct relationship between the power and the frequency, however, the turbine rotational speed and the frequency are disengaged. This allows for quicker changes of the VSHP output power by utilizing the rotational energy of the turbine and generator compared to a conventional power plant where the slow governor will limit the ancillary service capabilities. With that, new possibilities emerges as faster frequency control and other grid ancillary service, but it also necessitates proper co-ordination of the controls - and there will be new constraints that must be taken into account.

\begin{figure}[!t]
\centering
\begin{tikzpicture}[auto, node distance=1cm,>=latex']
    \path [draw=blue,snake it]
    (-1,0.5) -- (0,0.5) ;
    \path [draw=blue,snake it]
    (2.9,0.2) -- (3.3,0.2) ;
    \draw(0,0.7)--(0,0);
    \draw(-1,-1) .. controls (-0.4,-1) .. (0,-0.5);
    \draw(0,0)--(2.9,-0.3);
    \draw(0,-0.5)--(3,-0.8);
    \draw(3,-0.8)--(4.4,-3.8);
    \draw(3.3,-0.3)--(4.72,-3.42);
    \draw(2.9,-0.3)--(2.9,0.7);
    \draw(3.3,-0.3)--(3.3,0.7);
    \draw[red,thick](5,-4) circle (0.6);
    \draw(5.4,-4.4) .. controls (5.9,-5.0) .. (6.9,-4.0);
    \draw(4.8,-4.55) .. controls (5.9,-5.4) .. (7.2,-4.1);
    \draw(6.5,-3.6) .. controls (6.7,-3.9) .. (6.9,-4.0);
    \draw(7.2,-4.1)--(7.4,-4.1);
    \path [draw=blue,snake it]
    (6.6,-3.75) -- (7.4,-3.75) ;
    \node [align=center,rotate=-6] at (1.5,-0.4) {$q_{hr}\rightarrow$};
    \node [align=center,rotate=-6] at (1.5,-0.9) {Head race tunnel};
    \node [align=center,rotate=-6] at (1.5,0.1) {$f_{p2}$};
    \node [align=center,rotate=-67] at (4.25,-3) {$q\rightarrow$};
    \node [align=center,rotate=-67] at (3.5,-2.4) {Penstock};
    \node [align=center,rotate=-67] at (4.4,-2.0) {$f_{p1}$};
    \node [align=center] at (5,-4) {Turbine};
    \node [align=center] at (-0.4,-0.2) {Reservoir};
    \node [align=center] at (4.2,-0.2) {Surge tank};
    \node [align=center] at (2.6,0.1) {$f_{0}$};
    
    \node [align=center] at (6.8,-2.1) {$h_{st}$};
    \draw[->](6.8,-1.7)--(6.8,0.2);
    \draw[->](6.8,-2.5)--(6.8,-3.75);
    
    \node [align=center] at (7.2,-1.9) {$1$};
    \draw[->](7.2,-1.6)--(7.2,0.5);
    \draw[->](7.2,-2.2)--(7.2,-3.75);
    
    \draw[dotted](0,0.5)--(7.2,0.5);
    \draw[dotted](3.3,0.2)--(6.8,0.2);
    
    \node [align=center] at (6,-4) {$h$};
    \draw[->](6,-3.8)--(6,-3.4);
    \draw[->](6,-4.2)--(6,-4.6);
    
    \draw[dotted](5,-3.4)--(6,-3.4);
    \draw[dotted](5,-4.6)--(6,-4.6);
    
    \draw(3.7,-4) circle (0.6);
    \node [align=center] at (3.7,-4) {Gen};
    
    \draw[thick](4.4,-4)--(4.3,-4);
    
    \draw(0.6,-3.5)--(1.6,-3.5);
    \draw(0.6,-4.5)--(1.6,-4.5);
    \draw(0.6,-3.5)--(0.6,-4.5);
    \draw(1.6,-3.5)--(1.6,-4.5);
    \draw(0.6,-4.5)--(1.6,-3.5);
    \node [align=center] at (0.85,-3.75) {$\sim$};
    \node [align=center] at (1.35,-4.25) {$=$};
    
    \draw(3.1,-4)--(2.9,-4);
    
    \draw(1.9,-3.5)--(2.9,-3.5);
    \draw(1.9,-4.5)--(2.9,-4.5);
    \draw(1.9,-3.5)--(1.9,-4.5);
    \draw(2.9,-3.5)--(2.9,-4.5);
    \draw(1.9,-4.5)--(2.9,-3.5);
    \node [align=center] at (2.15,-3.75) {$=$};
    \node [align=center] at (2.65,-4.25) {$\sim$};
    
    \draw(1.9,-4)--(1.6,-4);
    
    \draw(-0.7,-3.5)--(0.3,-3.5);
    \draw(-0.7,-4.5)--(0.3,-4.5);
    \draw(-0.7,-3.5)--(-0.7,-4.5);
    \draw(0.3,-3.5)--(0.3,-4.5);
    \node [align=center] at (-0.2,-4) {Grid};

    \draw(0.3,-4)--(0.6,-4);
    
    \node [align=center] at (4.35,-4.7) {$\leftarrow P_{m}$};
    \node [align=center] at (0.45,-4.7) {$\leftarrow P_{e}$};
    \node [align=center] at (4.35,-5.1) {$\omega \Rcirclearrow$};
    \node [align=center] at (0.45,-5.1) {$f \sim$};
    
\end{tikzpicture}
\caption{Waterway layout} \label{figww}
\end{figure}
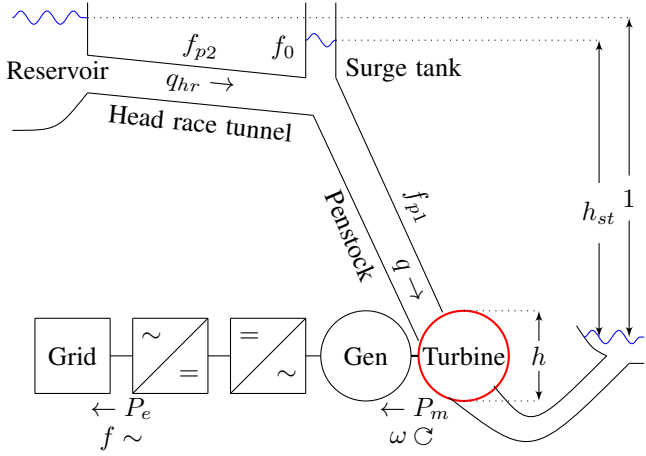




This paper is organized as follows: The MPC theory and the control objectives for the MPC controller are presented in Section \ref{MPC} while the development of the MPC model and Kalman filter are presented in, respectively, Sections \ref{MPCmodel} and \ref{Kalman}. The results and discussions are given in Section \ref{Results} and the conclusion in Section \ref{Conclusion}.

 






\section{Model Predictive Control}\label{MPC}

MPC controllers offer great advantages compared to transitionally PID controllers, although they are more complex. They are multiple-input, multiple-output (MIMO) controllers, they offer a faster and smoother response and lower rising time, settling time and overshoots compared to PID controllers and they are more robust. While the PID controller is a linear controller, MPC can handle non-linear systems as power electronics. However, a proper system model is needed for the design of the MPC controller.



MPC is a closed-loop optimization problem where a discrete-time model is optimized on a time horizon from $t=0$ to $t=N$. Only the inputs for the first time step are used and the optimization problem is recalculated for the next time step, with the new initial state values.  

A linear MPC model with quadratic objective function and linear constraints called an output feedback linear MPC, is used in this paper. The model \eqref{objfunk}-\eqref{posdef} includes cost for the error of state/variables values, changes in state values, the error of input values, changes in input values and cost for exceeding the constraints on the states with the use of slack variables.

\begin{multline}\label{objfunk}
    \min_{x \in \mathbb{R}^n, u \in \mathbb{R}^m} f(x,u) = \sum_{t=0}^{N-1} \frac{1}{2} x_{t+1}^{\text{T}}Q_{t+1}x_{t+1}  \\
    + d_{xt+1} x_{t+1} + \frac{1}{2} \Delta x_{t+1}^{\text{T}} Q_{\Delta t} \Delta x_{t+1}+ \frac{1}{2} u_{t}^{\text{T}}R_{t}u_{t}  \\ 
    + d_{ut} u_{t} + \frac{1}{2} \Delta u_{t}^{\text{T}} R_{\Delta t} \Delta u_{t} + \rho^{\text{T}} \epsilon + \frac{1}{2} \epsilon^{\text{T}}S \epsilon
\end{multline}


subjected to

\begin{equation}\label{constr}
    \begin{aligned}
        x_{t+1} &= A_t x_t + B_t u_t  \\
        x_0, u_{-1} &= \text{given} \\
        x^{\text{low}} - \epsilon &\leq x_t \leq x^{\text{high}} +\epsilon \\
        -\Delta x^{\text{high}} &\leq \Delta x_t \leq \Delta x^{\text{high}}\\
        A_{ineq} x_t &+ B_{ineq} u_t \leq b_{ineq} \\
        u^{\text{low}} &\leq u_t \leq u^{\text{high}} \\
        -\Delta u^{\text{high}} &\leq \Delta u_t \leq \Delta u^{\text{high}} \\
    \end{aligned}
    \quad
    \begin{aligned}
        t &= 0, \dots, N-1 \\
        \\
        t &= 1, \dots, N\\
        t &= 1, \dots, N\\
        t &= 1, \dots, N\\
        t &= 0, \dots, N-1\\
        t &= 0, \dots, N-1\\
    \end{aligned}
\end{equation}

where

\begin{equation}\label{posdef}
    \begin{aligned}
        Q_t &\succeq 0 \\
        Q_{\Delta t} &\succeq 0 \\
        R_t &\succeq 0 \\
        R_{\Delta t} &\succeq 0 \\
        \Delta x_t &= x_t -x_{t-1} \\
        \Delta u_t &= u_t -u_{t-1} \\
        z^{\text{T}} &= \left( x_1^{\text{T}}, \dots , x_N^{\text{T}},u_0^{\text{T}}, \dots , u_{N-1}^{\text{T}} \right) \\
        \epsilon &\in \mathbb{R}^n_x \geq 0 \\
        \rho &\in \mathbb{R}^n_x \geq 0 \\
        S &\in \text{diag} \left\{s_1, \dots, s_{n_x} \right\}, s_i \geq 0,
    \end{aligned}
    \begin{aligned}
        t &= 1, \dots, N\\
        t &= 1, \dots, N\\
        t &= 0, \dots, N-1\\
        t &= 0, \dots, N-1\\
        t &= 1, \dots, N\\
        t &= 0, \dots, N-1\\
        \\
        \\
        \\
          i&=\left\{1,\dots,n_x\right\}
    \end{aligned}
\end{equation}


The optimization problem is solved by the \texttt{quadprog} function in MATLAB.

\subsection{Control Objectives for the MPC Controller}\label{ch:objectives}

The MPC controller solves the optimization problem to find the optimal inputs $u$; the power reference $P_g^*$ and the guide vane reference $g^*$, while handling all constraints defined in the MPC model. The main tasks of the MPC in this paper are:

\begin{itemize}
    \item Primary frequency control:
    \begin{itemize}
        \item Provide power reference $P_g^*$ to the VSG.
        \item Minimize deviation in grid frequency $\Delta f$.
        \item Keep the converter power $P_g$ within its limits.
    \end{itemize}
    \item Hydraulic system control:
    \begin{itemize}
        \item Provide guide vane reference $g^*$ to the turbine.
        \item Minimize the operation of guide vane opening $g$ to reduce wear and tear.
        \item Minimize the rate of change of $g$ to reduce water hammering and mass oscillation.
        \item Keep the surge tank level $h_{st}$ within its limit and close to the stationary value.
        \item Keep the water flow $q$ above its minimum level.
        \item Optimize the rotational speed of the turbine $\omega$.
    \end{itemize}
    \item Turbine speed control:
    \begin{itemize}
        \item Keep the rotational speed of the turbine $\omega$ within the limits and close to its optimal speed.
        \item Make sure that $\omega$ will recover after a disturbance.
    \end{itemize}
\end{itemize}

Other possible tasks for the MPC, not implemented in this paper, will be:

\begin{itemize}
    \item Power oscillation damper (POD).
    \item Optimize the control of guide vane opening $g$ to minimize water hammering and mass oscillation.
    \item Voltage control.
\end{itemize}

Some of these control objectives are conflicting. For instance, fast regulation of the guide vane opening $g$ reduces the deviation in turbine rotational speed $\omega$, however, this will increase the deviation in the surge tank level $h_{st}$ and increase mass oscillation and water hammering. The cost of changing $g^*$, of deviations in $h_{st}$ and of exceeding the limits of $h_{st}$ will reduce the rate of change of $g$. Similarly, the cost of deviation in $\omega$ will increase the rate of change of $g$.

\section{MPC Dynamic Model}\label{MPCmodel}

This section presents the MPC model with its costs and constraints. Finally, linearization and discretization of the model are shown. 


The step length of the MPC model is set to $\Delta t = 0.2 s$ to cover the low frequency ($<0.5 Hz$) dynamics of the waterway system. An appropriate number of time steps is found to be $N=41$, resulting in a prediction horizon of $8.2s$. Based on simulation studies, we have found that the prediction horizon is long enough to ensure the performance and stability of the control system. Control input blocking is used to reduce the number of control input decision variables. The block sizes are equal to the step sizes for the first 10 steps, thereafter the sizes of the blocks gradually increase such that the total number of blocks becomes $m=21$. 

The MPC model is based on the models presented in \cite{reigstad2019modelling} and \cite{reigstad2019variable}, and is combined with the VSG presented in \cite{reigstad2019virtual}. These papers present all parameters and variables that are not explained in this paper. Sections \ref{sec:gov} to \ref{sec:gm} presents the differential-algebraic equations (DAE) \eqref{eqgov}-\eqref{eqgrid} of the MPC model. These are necessary to construct the matrices $A_t$ and $B_t$ in the equality constraints in \eqref{constr} as presented in Section \ref{sec:lin}. The inequality constraints of \eqref{constr} and the cost function \eqref{objfunk} are constructed from the information given in respectively Sections \ref{ch:limits} and \ref{ch:costs}.

\subsection{Governor}\label{sec:gov}



The governor can either set the rotational speed reference $\omega^*$ or the governor control can be performed by the MPC, setting the guide vane opening reference $g^*$. Although the open-loop system without a governor control is unstable, the latter alternative is chosen in this paper since the MPC will manage the governor control. The guide vane opening $g$ is found as

\begin{equation}\label{eqgov}
    \begin{split}
        \dot{g} &= \frac{1}{T_{G}} \left( g^* - g \right)
    \end{split}
\end{equation}

\subsection{Waterway}

The hydraulic system is modelled by the Euler turbine equation model presented in \cite{reigstad2019modelling}. To reduce the number of states, the penstock water column is assumed to be inelastic, and the differential equations for the waterway are thereby given as:

\begin{equation}\label{eqww}
    \begin{split}
        \dot{h}_{st} &= \frac{1}{C_s} \left( q_{\mathrm{hr}} -q\right)\\
        \dot{q}_{hr} &= \frac{1}{T_{\mathrm{w2}}} \left( 1 - h_{\mathrm{st}}+f_{0}{\left(q_{\mathrm{hr}}-q \right)}^2-f_{\mathrm{p2}}{q_{\mathrm{hr}}}^2 \right)\\ 
        h &= h_{st} - f_0 \left( q_{\mathrm{hr}} - q  \right)^2  - f_{\mathrm{p1}}  q  ^2\\
    \end{split}
\end{equation}

\subsection{Turbine}

The turbine model is based on the Euler turbine equation, as presented in \cite{nielsen2015simulation,reigstad2019modelling}.


\begin{equation}\label{eqturb}
    \begin{split}
        P_m &=   \frac{H_{\mathrm{Rt}}}{H_{\mathrm{R}}} \frac{Q_{R}}{Q_{Rt}} \\
        & \quad \left( \left( \frac{\xi q}{g}  \left( \tan{\alpha_{1R}} \sin{ \alpha_1}
         + \cos{\alpha_1} \right) \right) - \psi \omega \right) \frac{q \omega}{h}\\
        \alpha_1 &= \sin^{-1}{ \left(  \frac{Q_{\mathrm{R}}}{Q_{\mathrm{Rt}}} g \sin{\alpha_{1R} } \right)} \\
        \dot{q} &= \frac{1}{T_{\mathrm{w1}}} \left( h \frac{H_{\mathrm{R}}}{H_{\mathrm{Rt}}} -\sigma \left( \omega^2 -1  \right) -  \left( \frac{q }{g} \right)^2 \right) \frac{Q_{Rt}}{Q_{R}}\\
    \end{split}
\end{equation}

\subsection{Synchronous Generator}

To save simulation time, a simple first-order synchronous generator model \eqref{eqsg} is used in the MPC model. The torque must be used in the swing equation instead of the power since the rotational speed is not constant. Since the converter controller time constants are significantly smaller than the sampling time of the MPC, the electrical power of the synchronous generator is assumed to be equal to the output power of the VSHP $P_g$.

\begin{equation}\label{eqsg}
    \begin{split}
    \dot{\omega}&= \frac{1}{2H} \left( T_m - P_g/\omega - D \left( \omega^*-\omega \right) \right)\\
    \dot{\omega}&= \frac{1}{2H\omega} \left( P_m - P_g - D \left( \omega^*-\omega \right)\omega \right)\\
    \end{split}
\end{equation}







\subsection{Grid Converter}

To simplify the model, only the outer d-axis loop control of the grid converter, the active power control, is considered. This simplification is satisfactory since the inner controller is faster than the step length of the MPC and since the voltage control is not considered. The active power is controlled by a VSG, which is found to be more suitable for the purpose than the virtual synchronous machine (VSM) \cite{reigstad2019virtual}.

It is assumed that the converter output power $P_g$ equals the d-axis current $i_{g,d}$ such that

\begin{equation}\label{eqvsg}
    \begin{split}
        P_{g} &= i_{g,d} = k_{vsg,p} \Delta f +  k_{vsg,d} \Delta \dot{f}   + P_{g}^{*}\\
        \Delta f &= f^*-f\\
    \end{split}
\end{equation}

\subsection{Grid Model}\label{sec:gm}



The grid frequency is derived from the swing equation \cite{kundur1994power}.

\begin{equation}
\begin{split}\label{eqgrid}
    \Delta \dot{f} &= \frac{\omega_s}{2 H_g S_n} \left( P_g + P_{pb} - D_m \Delta  f \right)\\
\end{split}
\end{equation}

where $P_{pb}$ is the power balance of the grid without the VSHP; $P_{pb} = P_{generation} - P_{loads} - P_{losses}$. The values of the mean grid inertia $H_g=25.35p.u.$, the total rated power of all connected power producers $S_n=1p.u.$ and the damping of the grid $D_m = 0$ are assumed supplied from the TSO and are updated continuously.

The electrical power in the grid is estimated from the measured grid frequency $f$ and rate-of-change-of-frequency (ROCOF) $\dot{f}$ by the PLL.

\begin{equation}\label{eqPe}
\begin{split}
     P_{pb} &= -P_g + \frac{2H_g S_n}{\omega_s} \frac{\omega_{\dot{f}}}{s+\omega_{\dot{f}}} \Delta \dot{f} + D_m \frac{\omega_{f}}{s+\omega_{f}} \Delta f
\end{split}
\end{equation}

$f$ and $\dot{f}$ are filtered by first order filters with filter constants at respectively $\omega_{f}=0.625rad/s$ and $\omega_{\dot{f}}=0.25rad/s$. The total grid inertia $H_g$ and damping $D_m$ is

\subsection{Constraints and Slack Variables}\label{ch:limits}

The constraints on the inputs and variables $u$ are given in Table \ref{table_limits}. The guide vane opening reference $g^*$ is limited by the minimum and maximum values during normal operation and the converter power $P_g$ is limited by its maximal nominal power. Power transfer from the grid to the generator is blocked by setting the lower constraint of $P_g$ to zero. In addition, the change in $g^*$ from one step to the next is limited to $\Delta g^*_{max} = 0.2  \Delta t = 0.04$, which correspond to the maximum operational speed of the guide vane.

\begin{table}[!t]
\renewcommand{\arraystretch}{1.3}
\caption{Constraints on inputs and variables}
\label{table_limits}
\centering
\begin{tabularx}{1\linewidth}{|X|cc|}
    \hline
    Input & Min. value & Max. value \\
    \hline
    Guide vane opening reference $g^*$ & 0.1 & 1.3 \\
    Converter power $P_g$ & 0 & 1\\
    \hline
\end{tabularx}
\end{table}

\begin{table}[!t]
\renewcommand{\arraystretch}{1.3}
\caption{Slack variables}
\label{table_slack}
\centering
\begin{tabularx}{1\linewidth}{|X|ccc|}
    \hline
    Slack variable & Min. limit & Max. limit & Cost factor $S(i,i)$\\
    \hline
    Water flow $q$ & 0.3 & 1.3 & 1\\
    Surge tank level  $h_{st}$ & 0.5 & 1.1 & 1e6\\
    Turbine rot. speed $\omega$ & 0.7 & 2 & 1e5\\
    \hline
\end{tabularx}
\end{table}

To avoid non-convergence, slack variables are used instead of constraints on the state variables, as given in Table \ref{table_slack}. The turbine needs a minimum and maximum water flow $q$ to function properly, and a slack variable is used to add costs to the cost function if $q$ is outside its constraints. The next slack variable ensures that the surge tank level $h_{st}$ will be limited to the maximum pressure over the turbine, normally 1.1-1.15 p.u., or the maximum head of the surge tank. Exceeding these values may cause damage to the turbine blades or water to blow out of the surge shaft. This slack variable also avoids the surge tank level from becoming too low. Normally a sand trap is located between the surge shaft and pressure shaft. Too low surge tank level will cause sand to raise here and to be sent through the turbine, causing increased wear and tear and reduced lifetime of the turbine.

The third slack variable is related to the turbine rotational speed $\omega$, which is limited by the maximal rated speed of the generator. If this speed is exceeded, there is a high consequence risk of the poles to falling off. 

When $\omega$ is reduced and the converter output power $P_g$ is kept constant, the electrical torque will increase. The increase in mechanical torque will be less, and the MPC controller has to increase the guide vane opening $g$ to regain the reference turbine speed $\omega^*$. If $\omega$ decreases too much, the MPC controller will not be able to regain the reference turbine speed without reducing the converter output power $P_g$. A lower limit slack variable is therefore used on $\omega$ to prevent this situation.





\subsection{Costs in MPC Cost Function}\label{ch:costs}

The cost function includes costs for deviation in the grid frequency $\Delta f$, turbine rotational speed $\omega$ and the VSHP power reference $P_g^*$ from their reference value, as given in Table \ref{table_costs}. The costs for exceeding the constraints of the slack variables, given in Table \ref{table_slack}, are also included in the cost function.

\begin{table}[!t]
\renewcommand{\arraystretch}{1.3}
\caption{Cost on deviations in states and inputs}
\label{table_costs}
\centering
\begin{tabularx}{1\linewidth}{|X|cc|}
    \hline
    State/input & Reference value & Cost factor $Q(i,i)$\\
    \hline
    Grid frequency $\Delta f$& 0 & 0.01\\
    Turbine rotational speed $\omega$& $f(P_{pb})$, \eqref{eqturbopt} & 100\\
    VSHP power reference $P_g^*$ & 0.8 & 1000\\
    \hline
\end{tabularx}
\end{table}

The relative values of the costs determine how the MPC priorities between its objectives given in Section \ref{ch:objectives}. A high cost related to an objective causes the MPC controller to prioritize this objective to reduce the cost function. The objectives are prioritized as follows:

\begin{enumerate}
    \item Keep the surge tank level $h_{st}$ within its constraints to avoid damage of the hydraulic system.
    \item Keep the turbine rotational speed $\omega$ within its constraints to avoid undesired operation conditions of the hydraulic system and damage of the generator.
    \item Minimize the deviation in the VSHP power reference $P_g^*$ to assure that the VSHP is contributing to the frequency regulation as intended by the VSG.
    \item Minimize the deviation of the turbine rotational speed $\omega$ from the best efficiency operating point to increase the efficiency of the system.
    \item Keep the water flow $q$ within its constraints to avoid undesired operation conditions of the hydraulic system.
    \item Minimize the deviation in grid frequency $\Delta f$.
\end{enumerate}

The cost of deviation in $\Delta f$ is low and the cost of deviation in $P_g^*$ is high since the grid frequency control should primarily be performed by the VSG. The VSHP power reference $P_g^*$ is not supposed to compensate for deviations in the turbine rotational speed $\omega$ unless $\omega$ is predicted to go outside its constraints. The cost of deviations in $P_g^*$ is, therefore, higher than the cost of deviation in $\omega$. The deviations in $\omega$ will, when possible, be compensated only by adjusting the guide vane opening reference $g^*$ and thereby the mechanical power. However, if constraints on the surge tank level $h_{st}$, the water flow $q$ or the rate of change of the guide vane opening reference $\Delta g^*$ block the turbine rotational speed $\omega$ from being recovered within its limit, the VSHP power reference $P_g^*$ will be adjusted. In this way, situations, where the turbine rotational speed is reduced too much to be able to produce enough torque to increase again will be avoided.

\subsection{Reference Turbine Rotational Speed}

The optimal turbine rotational speed $\omega$ depends on the flow $q$ and thereby by the produced power. This is implemented in the MPC by letting the turbine rotational speed reference $\omega^*$ be a function of the VSHP output power $P_{g}$, as given in \eqref{eqturbopt}. The curve is based on the measured optimal speed of a reversible pump-turbine presented in \cite{iliev2019variable}. 

\begin{equation}\label{eqturbopt}
    \begin{split}
        0.85 < P_{g} \quad \quad \quad \quad \omega^* &= 1 +0.6 (P_{g}-0.85) \\
        0.73 < P_{g} < 0.85 \quad \omega^* &= 1 + 0.3(P_{g}-0.85) \\
          P_{g} < 0.73 \quad \omega^* &= 0.964 + 0.15(P_{g}-0.73) \\
    \end{split}
\end{equation}

\subsection{Linearization and Discretization of the Model}\label{sec:lin}

The system DAEs are given from \eqref{eqgov}, \eqref{eqww}, \eqref{eqsg}, \eqref{eqvsg} and \eqref{eqgrid} where

\begin{equation}
\begin{split}
    \dot{x} &= f(x,u) = [\Delta \dot{f}  \quad \dot{g}  \quad \dot{q} \quad \dot{q_{hr}} \quad \dot{h_{st}} \quad \dot{\omega}]^T \\
     x &= [\Delta f  \quad g  \quad q \quad q_{hr} \quad h_{st} \quad \omega]^T\\
     u &= [P_g^* \quad  P_{pb} \quad  g^*]^T
\end{split}
\end{equation}

The stationary operation point $x_s$ is found from the previous estimation of the grid power balance $P_{pb}$ and the previous value of the VSHP power reference $P_g^*$ by solving the equation $\dot{x}_s = 0$ for $g^*=g$. The system is linearized around this point as given by \eqref{lin}.

\newcommand{\at}[2][]{#1|_{#2}}
\begin{equation}\label{lin}
\begin{split}
     \Delta \dot{x} &= A_c \Delta x + B_c \Delta u \\
     \Delta \dot{y} &= C_c \Delta x + D_c \Delta u \\
     A_c &= \frac{\delta f}{\delta x}\at[\bigg]{\left( x_s, u_s \right)} = \begin{bmatrix} 
        \frac{\delta f_1}{\delta x_1}\at[\Big]{\left( x_s, u_s \right)} & \dots &    \frac{\delta f_1}{\delta x_n}\at[\Big]{\left( x_s, u_s \right)}    \\
        \vdots & \ddots & \vdots    \\
        \frac{\delta f_n}{\delta x_1}\at[\Big]{\left( x_s, u_s \right)} & \dots &    \frac{\delta f_n}{\delta x_n}\at[\Big]{\left( x_s, u_s \right)}    \\
        \end{bmatrix} \\
    B_c &= \frac{\delta f}{\delta u}\at[\bigg]{\left( x_s, u_s \right)} = \begin{bmatrix} 
    \frac{\delta f_1}{\delta u_1}\at[\Big]{\left( x_s, u_s \right)} & \dots &    \frac{\delta f_1}{\delta u_n}\at[\Big]{\left( x_s, u_s \right)}    \\
    \vdots & \ddots & \vdots    \\
    \frac{\delta f_n}{\delta u_1}\at[\Big]{\left( x_s, u_s \right)} & \dots &    \frac{\delta f_n}{\delta u_n}\at[\Big]{\left( x_s, u_s \right)}    \\
    \end{bmatrix} \\
\end{split}
\end{equation}

where $\Delta x = x- x_s$ and $\Delta u = u- u_s$ are the errors from the linearization point.

Next, the model is discretized as shown in \eqref{dis}, where $\Delta t$ is the step time length.

\begin{equation} \label{dis}
\begin{split}
    A_t &= A_c \Delta t + I\\
    B_t &= B_c \Delta t \\
\end{split}
\end{equation}

For each time step, a new stationary operation point based on the previous inputs and a new linearized function are found, and the equality constraints are updated with the new state system matrices. Cost matrices and inequality constraints must also be updated according to the new linearization point. 

\begin{figure}[!htb]
\centering
\begin{tikzpicture}[auto, node distance=1cm,>=latex']

    \node [block] (Pe) {\begin{tabular}{c}
         Estimate $P_{pb}$ \\
         \eqref{eqPe} 
    \end{tabular}};
    \node [block, below of=Pe, node distance =2cm] (x0) {\begin{tabular}{c}
         Calculate stationary values \\
         $\dot{x}_s=0$\\
         $g^* = g$\\
         $u_s =  [P_g^* \quad  P_{pb} \quad  g^*]^T$
    \end{tabular}};
    \draw [->] (Pe) -- node[anchor=west,pos=0.5]{$P_{pb}$} (x0);
    
    \node [block, below of=x0, node distance =2.5cm] (li) {\begin{tabular}{c}
         Linearize \eqref{lin} \\
         Update inequality \\ 
         constraints
    \end{tabular}};
    \draw [->] (x0) -- node[anchor=west,pos=0.5]{$x_s,u_s$} (li);
    \node [block, below of=li, node distance =2.0cm] (quad) {\begin{tabular}{c}
         Solve optimization \\
         problem \eqref{objfunk}
    \end{tabular}};
    \draw [->] (li) -- node[anchor=west,pos=0.5]{} (quad);
    
    \node [block, right of=li, node distance =4.0cm] (delt) {\begin{tabular}{c}
         $x_0 = \hat{x}_{kf}- x_s$ 
    \end{tabular}};
    \draw [->] (delt) |- node[anchor=west,pos=0.2]{$x_0$} (quad);
    
    \node [block, above of=delt, node distance =2.5cm] (kalman) {\begin{tabular}{c}
         Kalman filter \\
         \eqref{eqkalm} - \eqref{eqkalm4} 
    \end{tabular}};
    \draw [->] (kalman) -- node[anchor=west,pos=0.5]{$\hat{x}_{kf}$} (delt);
    
    \node [block, above of=kalman, node distance =2cm] (sys) {\begin{tabular}{c}
         Power\\
         system \\
         \cite{reigstad2019modelling,reigstad2019variable,reigstad2019virtual} 
    \end{tabular}};
    \draw [->] (sys) -- node[anchor=west,pos=0.5]{$y_{kf},\omega$} (kalman);
    \draw [->] (sys) -- node[anchor=south,pos=0.5]{$\Delta f,\Delta \dot{\omega}_g$} (Pe);
    \draw [->] (delt) -- node[anchor=south,pos=0.5]{$x_0$} (li);
    \draw [->] (x0) -- node[anchor=south,pos=0.5]{$x_s$} (delt);
    \node [point, below of =quad,node distance=1.2cm] (quadb){};
    \node [point, right of =quadb,node distance=5.9cm] (quadbr){};
    \draw [->] (quad) -- node[anchor=west,pos=0.5]{$g^*,P_g^*$} (quadb) --  (quadbr) |- node[anchor=south,pos=0.75]{$g^*$,$P_g^*$} (sys);
    \node [point, above of =quadbr,node distance=5.7cm] (quadbra){};
    \draw [->] (quadbra) --  node[anchor=south,pos=0.5]{$g^*$}(kalman);
    \node [point, left of =quadb,node distance=2.7cm] (quadbl){};
    \draw [->] (quad) -- (quadb) --  (quadbl) |- node[anchor=south,pos=0.7]{$P_g^*$} (x0);

\end{tikzpicture}
\caption{Float diagram for MPC controller} \label{figFloat}
\end{figure}
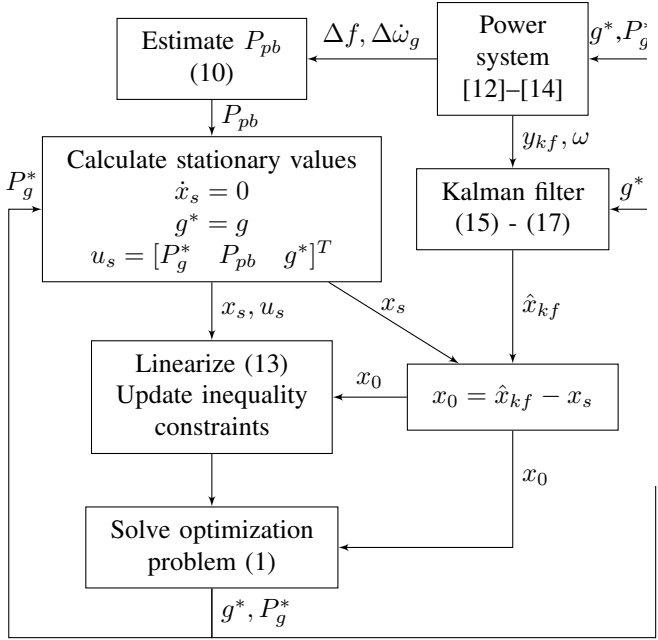

The steps of the MPC are explained in Figure \ref{figFloat}. The VSHP inputs $g^*$ and $P_g^*$ from the previous solution of the optimization problem are applied to the power system. At the next time step, the grid power balance $P_{pb}$ is estimated to calculate the stationary state values by setting $\dot{x}_s=0$. In parallel, the Kalman filter, explained in the next section, estimates the state values and the deviations from the stationary values are found. The system DAEs are then linearized based on the stationary values and cost matrices, and the inequality constraints are updated. Finally, the optimization problem is solved and the first inputs to the power system are found and applied.

\section{Kalman Filter}\label{Kalman}

A Continuous-Time Kalman filter is used to estimate the unmeasured variables in the hydraulic system. The guide vane opening $g$, the surge tank height $h_{st}$, the height over the turbine $h$ and the mechanical power $P_m$ are measured. The Kalman filter is designed to filter $g$ and $h_{st}$ and estimate values of the pressure tunnel flow $q$ and the headrace tunnel flow $q_{hr}$. The estimated values will be used as input to the MPC. The dynamical system model is:

\begin{equation} \label{eqkalm}
\begin{split}
    \dot{x}_{kf} &= A_{kf}x_{kf} +B_{kf}u_{kf} +G_{kf}w \\
    y_{kf} &= C_{kf}x_{kf} + D_{kf}u_{kf} +H_{kf}w +v  \\
\end{split}
\end{equation}

where

\begin{equation} \label{eqkalm2}
\begin{split}
    x_{kf} &= \left[ g \quad q \quad q_{hr} \quad h_{st}  \right]^T \\ 
    y_{kf} &= \left[ g \quad h_{st} \quad h \quad P_m \right]^T \\
    u_{kf} &= \left[ g^* \quad \omega \right]^T
\end{split}
\end{equation}

 The matrices $A_{kf}$, $B_{kf}$, $C_{kf}$ and $ D_{kf}$ are found by linearizing the hydraulic system model \eqref{eqgov} - \eqref{eqturb} at the initial stationary operation point. $w$ and $v$ are, respectively, white process noise and measurement noise. 
 


The Kalman filter equations are given as:
 
\begin{equation} \label{eqkalm4}
\begin{split}
    \dot{\hat{x}}_{kf} &= A_{kf} \hat{x}_{kf} + B_{kf} u_{kf} \\
    & \quad + L_{kf} \left( y_{kf} - C_{kf} \hat{x}_{kf} - D_{kf} u_{kf} \right) \\
    \begin{bmatrix}
        \hat{y}_{kf} \\
        \hat{x}_{kf}
    \end{bmatrix}
    &= 
    \begin{bmatrix}
        C_{kf} \\
        I
    \end{bmatrix}
    \hat{x}_{kf} + 
    \begin{bmatrix}
        D_{kf} \\
        0
    \end{bmatrix}
    u_{kf} 
\end{split}
\end{equation}

where the filter gain $L_{kf}$ is solved by an algebraic Riccati equation in MatLab \cite{lewis2017optimal, MATLAB:2018b}.


\section{Results and Discussion}\label{Results}




The dynamic performance of the MPC controller is tested on the grid presented in \cite{reigstad2019variable}. Cases with both overproduction and underproduction are investigated by first reducing the load by 160 MVA at Bus 7 at time $t=0s$ and thereby increasing the load back to the initial value at $t=60s$. 

\begin{figure}[!t]
\centering
    \includegraphics[scale=0.8]{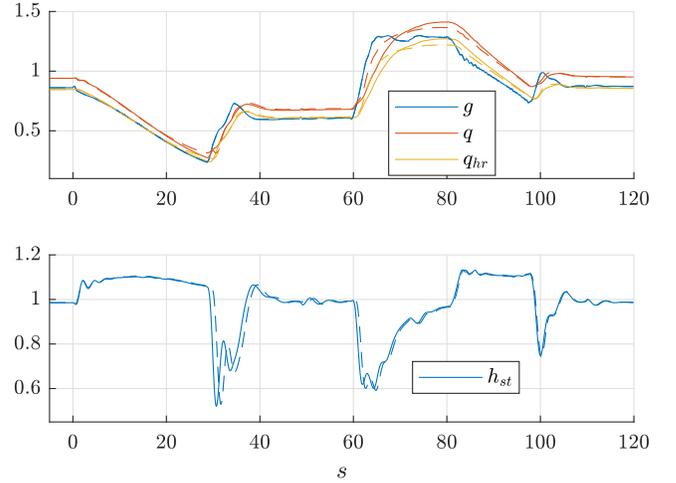}
    \caption{Performance of Kalman filter: Real values (solid) and estimations by the Kalman filter (dashed)} \label{fig612}
\end{figure}

Figure \ref{fig612} compares the real values of four states with the values estimated by the Kalman filter. The estimation of the guide vane opening $g$ is almost perfect since the reference value ($g^*$) is known. A small delay is observed for the other states; the turbine flow $q$, the headrace tunnel flow $q_{hr}$ and the surge tank head $h_{st}$.

Three different scenarios are investigated to show how the parameters of the MPC and VSG affect the grid and the hydraulic system:

\begin{enumerate}
    \item MPC: Initial settings, VSG: 1\% droop
    \item MPC: Initial settings, VSG: 4\% droop
    \item MPC: Turbine speed constraints reduced to $0.85-1.10 p.u.$, VSG: 1\% droop
\end{enumerate}

\begin{figure}[!t]
\centering
    \includegraphics[scale=0.79]{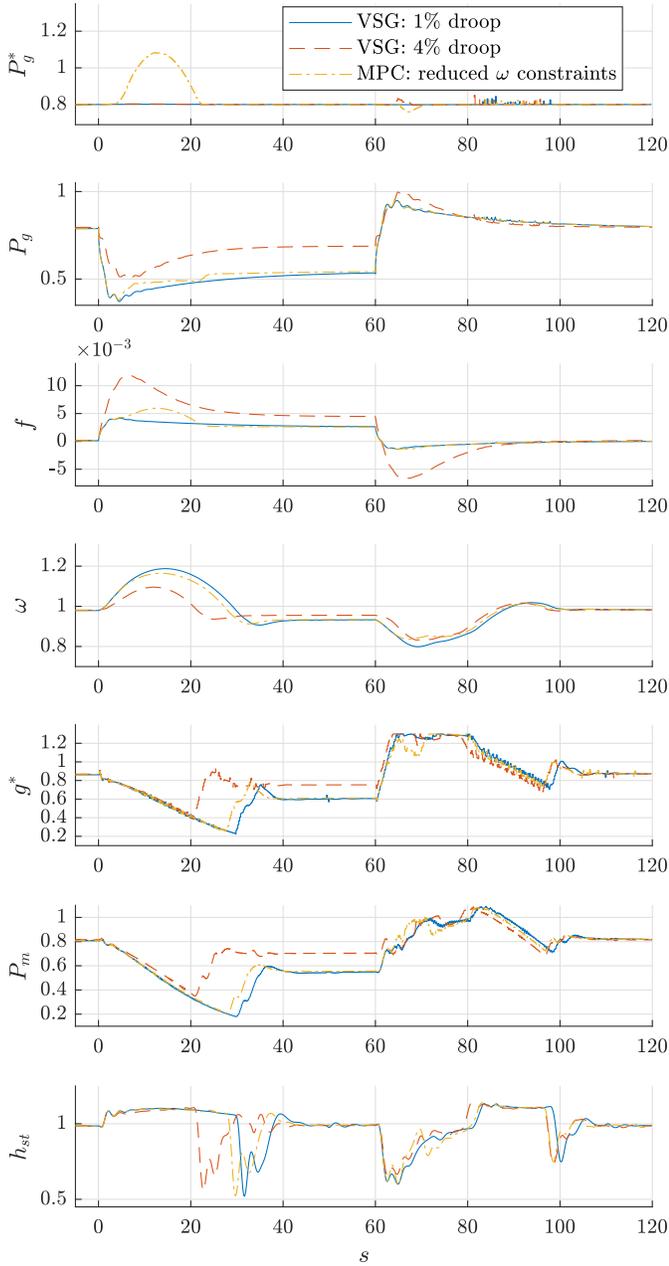}
    \caption{Dynamic performance at 1\% droop, 4\% droop and reduced limits on turbine rotational speed} \label{fig611}
\end{figure}

Figure \ref{fig611} shows the reference and the measured VSHP power $P_g^* , P_g$, the  grid frequency $f$, the turbine rotational speed $\omega$, the guide vane opening reference $g^*$, the turbine mechanical power $P_m$ and the surge tank head $h_{st}$. When the grid load is reduced at $t=0s$, the grid frequency $f$ immediately starts increasing because of overproduction in the system. The VSG reduces the VSHP output power $P_g$ depending on the droop; if the droop is low (1\%), $P_g$ is reduced by approximately 0.4 p.u. within 2 sec, and the peak frequency is limited to $0.4\%$. In this case, most of the loss reduction is actually compensated by the VSHP. With $4\%$ droop, the decrease in $P_g$ is less, causing a three times higher frequency deviation. 

The MPC minimizes its cost given in Section \ref{ch:costs} while fulfilling the constraints in Section \ref{ch:limits}. To reduce the cost of deviation in turbine rotational speed $\omega$, the MPC reduces the guide vane opening reference $g^*$ immediately to regain $\omega$ as fast as possible. However, the maximal $g^*$ step size is limited to the maximal operational speed of the governor. This constraint is active for the first time steps after the load reduction. The fast reduction in guide vane opening $g$ causes the surge tank head $h_{st}$ to increase close to its maximal value. To avoid $h_{st}$ from exceeding its maximal value, the MPC reduces the rate of change of $g^*$ and $g$ after 0.6 sec.

The guide vane opening $g$ is reduced as fast as possible until the turbine rotational speed $\omega$ is almost regained to its optimal value. Subsequently, $g$ increases. Since there is a larger deviation between the stationary value and the lower constraint of $h_{st}$ than of the stationary value and the higher constraint of $h_{st}$, $g^*$ and $g$ are allowed to increase faster than it decreases. Partly, the rate of change of the guide vane opening is limited by the maximum step size of $g^*$.

After 60 sec, the grid load increases by 160 MW, back to its initial value. This causes the grid frequency $f$ to drop. The guide vane opening reference $g^*$ increases with its maximal rate of change until it almost reaches its maximum value. The maximal deviation in turbine rotational speed $\omega$ is less for the case of load increase compared to the case of load decrease. The rate of change of the guide vane is faster since the lower constraint of the surge tank head $h_{st}$ is not active for most of the time. Thereby, the turbine mechanical power $P_m$ changes faster to recover $\omega$. This is a very important quality of the proposed MPC control since too low rotational speed must be avoided. In cases with high VSHP output power $P_g$ and low turbine rotational speed $\omega$, the turbine might not be able to deliver enough power to regain $\omega$ without reducing $P_g$. If $P_g$ is not reduced in this case, the turbine stops. While a conventional governor control increases and decreases the guide vane opening $g$ at the same speed, the MPC controller makes it possible to increase the opening speed of $g$. This reduces the minimum rotational speed, and thereby avoid situations where $P_g$ has to be reduced to regain $\omega$.

The third case in Figure \ref{fig611} shows how the MPC handles situations where both surge tank height $h_{st}$ and the turbine rotational speed $\omega$ exceed its constraints. In this case, the constraints of $\omega$ are reduced to $0.85-1.10 p.u.$. At $t=10s$, $h_{st}$ has reached its maximal value and limits the rate of change of guide vane reference $g^*$. It is therefore not possible to close $g$ faster to reduce $\omega$, which is simultaneously getting close to its maximal value. Since the cost of the $h_{st}$ and $\omega$ slack variables are higher than the cost of deviations in VSHP output power reference $P_g^*$, the MPC increases $P_g^*$ to avoid $h_{st}$ and $\omega$ from exceeding its constraints. This causes a temporary increase in VSHP output power $P_g$ and grid frequency $\Delta f$.

\begin{figure}[!t]
\centering
    \includegraphics[scale=0.8]{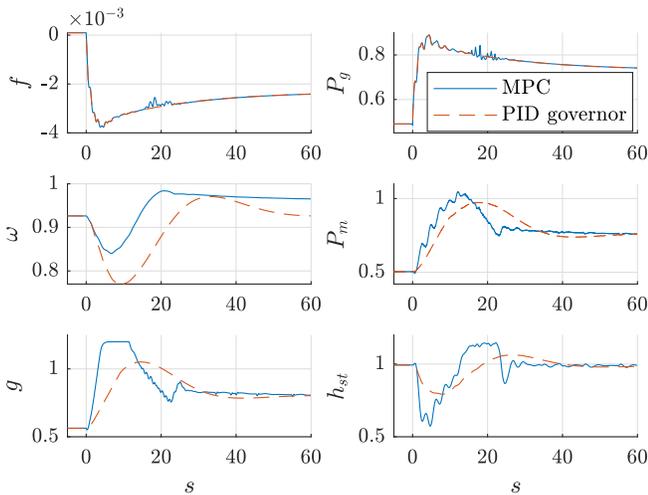}
    \caption{Dynamic performance after generator loss; with MPC (blue) and PID governor from \cite{reigstad2019variable} (red). VSG with 1\% droop is utilized in both cases.} \label{fig622}
\end{figure}

The performance of the controller system after disconnection of half of the generators at G2 at $t=0$ is shown in Figure \ref{fig622}. To illustrate its benefits, the MPC controller is compared to the governor control presented in \cite{reigstad2019variable}, however, the VSG with 1 \% droop controls the grid converter output power. Since the MPC considers the limitations in surge tank level $h_{st}$, the guide vane opening $g$ can be increased faster until its maximum value is reached or the minimum value of $h_{st}$ is reached. This results in higher turbine mechanical power $P_m$ and thereby less deviation in turbine rotational speed $\omega$ and higher efficiency of the turbine. The more aggressive control of the guide vane opening $g$ causes higher deviation and more oscillations in the surge thank level $h_{st}$, however, this can be tolerated since the MPC controller handles the system constraints. Due to the increased performance of the turbine control and lower deviation in turbine rotational speed $\omega$, it is possible to increase the FFR delivery.

The step responses in Figures \ref{fig612}, \ref{fig611} and \ref{fig622} show that the linearized MPC model is not a perfect model of the system. For instance, the surge tank head $h_{st}$ should be closer to its maximum constraint between $0-30s$ and the overshoots in turbine rotational speed $\omega$ and guide vane opening $g$ should be less. The use of a nonlinear MPC will improve the calculation of the turbine torque and thereby increase the precision of the control and reduce or eliminate these problems.

\section{Conclusion}\label{Conclusion}

As the share of wind and solar energy production increases, more flexible production and loads are required to control the balance of the grid in order to maintain the power system security. By utilizing the rotational energy of the turbine and the generator, VSHPs are able to deliver both VI and FFR. However, an advanced MIMO control system is needed to optimize the control and to ensure that the hydraulic and electric variables are within their constraints. A control system with an overall MPC and VSG control of the grid-connected converter is developed to fulfil the control objectives. When a grid frequency deviation occurs, the VSG controls the output power of the converter to reduce the frequency deviation. Thereby, the MPC will primarily control the turbine guide vane opening to regain the nominal turbine rotational speed. The speed of the control will be faster than for a conventional governor control since the MPC maximizes the rate of change of the guide vane opening while considering the surge tank head guide vane speed constraints. In cases where the turbine rotational speed could not be kept within its limits due to these constraints, the MPC will adjust the VSG power reference and thereby change the VSHP output power to regain the turbine rotational speed. The linearization of the MPC model causes inaccurate prediction and overshoots that may be improved by the use of nonlinear MPC.

\bibliographystyle{IEEEtran}
\bibliography{Paper5}

\end{document}